\documentclass[fleqn,usenatbib]{mnras}

\usepackage{newtxtext,newtxmath}

\usepackage[T1]{fontenc}
\usepackage{ae,aecompl}

\usepackage{graphicx}	
\usepackage{amsmath}	
\usepackage{amssymb}	
\usepackage{subfig}

\title[Nonlinear variability in GX~339-4]{Nonlinear Variability of Quasi-Periodic Oscillations in GX~339-4}

\author[Arur \& Maccarone.]{
K. Arur,$^{1}$\thanks{E-mail: kavitha.arur@ttu.edu}
T. J. Maccarone $^{1}$
\\
$^{1}$Department of Physics and Astronomy, Texas Tech University,Lubbock,TX,79409-1051,USA
}

\date{Accepted XXX. Received YYY; in original form ZZZ}

\pubyear{2018}

\begin{document}
\label{firstpage}
\pagerange{\pageref{firstpage}--\pageref{lastpage}}
\maketitle

\begin{abstract}
We examine the nonlinear variability of quasi-periodic oscillations (QPOs) using the bicoherence, a measure of phase coupling at different Fourier frequencies. We analyse several observations of RXTE/PCA archival data of the black hole binary GX 339-4 which show QPOs. In the type C QPOs, we confirm the presence of the `hypotenuse' pattern where there is nonlinear coupling between low frequencies that sum to the QPO frequency as well as the `web' pattern where in addition to the hypotenuse, nonlinear coupling between the QPO frequency and the broadband noise is present. We find that type B QPOs show a previously unknown pattern. We also show that the bicoherence pattern changes gradually from `web' to `hypotenuse' as the source moves from a hard intermediate state to a soft intermediate state. Additionally, we reconstruct the QPO waveforms from six observations using the biphase. Finally, we present a scenario by which a moderate increase in the optical depth of the hard X-ray emission region can explain the changes in the non-linearity seen during this transition.
\end{abstract}

\begin{keywords}
stars: accretion, accretion discs -- individual: GX 339-4 -- methods: statistical -- binaries: close
\end{keywords}



\section{Introduction}

Quasi-periodic oscillations (QPOs) are features that appear as modest-width peaks (i.e. with quality factors \footnote{Ratio between QPO frequency and the FWHM of the QPO peak} of $\sim$few to $\sim$10) in the power spectra of X-ray binaries, and are considered to be excellent probes of the regions closest to the compact objects. Low frequency QPOs (LFQPOs), are seen at frequencies between $\sim$0.1Hz and $\sim$10Hz \citep[see eg.][for an overview]{Belloni2011}. Based on their amplitudes, quality factors, and other aspects of the power spectra, these LFQPOs, originally classified into 3 types: A, B and C in XTE~J1550-564 \citep{Wijnands1999,Remillard2002}, have since been observed in several black hole binaries (BHB).  

Black hole binaries also show variability on longer time scales, leading to with spectral state transitions, where the spectra change from a power-law dominated hard state, to a soft state that is dominated by a strong disk blackbody component \citep{Tananbaum1972}. These changes can be clearly illustrated using the Hardness Intensity Diagram (HID) \citep{Miyamoto1999, Maccarone2003, Homan2001,Homan2005,Belloni2005}, where different portions of the q-shaped track correspond to the different states of the BHB. The four main states are the Low/Hard state (LHS), the Hard Intermediate State (HIMS), the Soft Intermediate State (SIMS) and the High/Soft state (HSS) \citep{Belloni2005}. In addition to the changes seen in the HID, detailed timing and spectroscopic analyses \citep[e.g][]{Kalemci2013} are essential for a complete understanding of the underlying processes. The association of LFQPOs with the different spectral states suggests that understanding the origin of QPOs is fundamental in understanding these different states.  

Several different models have been proposed to explain the origin of these QPOs, such as the Lense-Thirring precession of the accretion disk \citep{Stella1998, Ingram2011}, spiral structures in the accretion disk \citep{Varniere2017} and instabilities in the accretion disk \citep{Abramowicz1995}. While many of the models can reproduce the frequencies of the peaks in the power spectra, there is still no consensus about the mechanism by which these QPOs are produced. Thus it is important to go beyond the power spectrum, and study the non-linear variability present in the different states. 

Recently, attempts have been made to understand the non-linear variability in X-ray binaries.  Nonlinearity has been shown by presence of a linear rms-flux relation \citep{Uttley2001}  and has also been seen in the hard state of Cygnus X-1 using the bispectrum and bicoherence, which are relative measures of phase couplings among three different Fourier components \citep{Maccarone2002}. Much of the phenomenology of X-ray binary variability is seen in active galactic nuclei, cataclysmic variables and young stellar objects, suggesting that most of the basic physics is generic to the process of accretion \citep{Scaringi2015}.  In GRS 1915+105, \cite{Maccarone2011} found that there was a peak in the bicoherence wherever harmonics of the QPO were present. It has been shown that the bispectrum can be used to distinguish between different models that produce very similar power spectra \citep{Maccarone2005}.

In this paper we use the bicoherence to examine the non-linear variability of multiple QPOs observed in GX 339-4 using the Rossi X-ray Timing Explorer (RXTE). In Section~\ref{sec:bispectrum} we give an introduction to the bispectrum, followed by a brief overview of the data in Section~\ref{sec:data}. In Section~\ref{sec:results} we present the results of our analysis and show that the type of variability changes as the source moves from the LHS/HIMS to the SIMS as well as reconstructed waveforms for a subset of the observations. We then propose a  scenario to explain the observed results in Section~\ref{sec:discussion} before presenting our conclusions.

\section{Bicoherence}
\label{sec:bispectrum}

The bispectrum is a higher order Fourier spectral analysis technique that can be used to study the non linear properties of the time series \citep{Tukey1953}. The bispectrum of a time series is given by:

\begin{equation}
    B(k,l) =\frac{1}{K} \sum_{i=0}^{K-1}X_i(k)X_i(l)X^*_i(k+l)
	\label{eq:bispectrum}
\end{equation}

where K is the number of segments of the time series, $X_i(f)$ is the Fourier transform of the $i$th segment of the time series at frequency f, and $X^*_i(f)$ is the complex conjugate of $X_i(f)$. The real component of the bispectrum describes the skewness of the flux distribution of the light curve, while the imaginary component describes the reversibility of the time series in a statistical sense. The bispectrum is a complex number that can be represented using an amplitude and a phase. This phase, which is called the biphase, thus holds information about the shape of the underlying lightcurve. For a more detailed discussion of the biphase, see \cite{Maccarone2013}.

A related term is the bicoherence, which is analogous to the cross-coherence function that is more familiar to most X-ray astronomers \citep{Nowak1996}. The bicoherence has a value between 0 and 1, where 0 indicates that there is no non-linear coupling between the phases of the different Fourier frequencies, and 1 indicates total coupling.  The bicoherence is given by

\begin{equation}
    b^2 =\frac{\left|\sum X_i(k)X_i(l)X^*_i(k+l) \right|^2}
    {\sum|X_i(k)X_i(l)|^2 \sum |X_i(k+l)|^2} 
	\label{eq:bicoherence}
\end{equation}

using the normalization proposed by \cite{Kim1979}. Here, $b^2$ measures the fraction of power at the frequency $k+l$ due to the coupling of the three frequencies.  

As the bicoherence is forced to lie between 0 and 1, it will have a non-zero mean arising from errors even when there is no phase coupling is present ~\citep{Maccarone2002}. Thus a bias of 1/K is subtracted from any bicoherence measurements.

\section{Data Overview}
\label{sec:data}

GX~339-4 is a Low Mass X-ray Binary (LMXB) that consists of a  black hole of mass 2.3$M_{\odot}$ < M$_{BH}$ < 9.5$M_{\odot}$ \citep{Heida2017}, and has undergone multiple X-ray outbursts since its discovery \citep{Markert1973}. In this paper we analyze RXTE archival observations of GX 339-4 during its outbursts in 2002, 2004, 2007 and 2010 where QPOs were detected by \cite{Motta2011}. 

The data were obtained in several modes, and for our analysis we used \textsc{Single Bit, Event} and \textsc{Good Xenon} mode data. The data were filtered to exclude periods of high offset, Earth occultation and passage through the South Atlantic Anomaly (SAA). Data above 30 keV are excluded, since for the softer states the hard X-rays are mostly contributed by the background. The data were then binned with a time resolution of 1/128s, and Fourier transformed using 16s long segments of the data. The Fourier transformed data were averaged and normalized according to \cite{Leahy1983} to produce the power spectra.  The patterns are seen most clearly in the bicoherence plots of observations that are both long (>1ks) and have a high count rate (>750 counts/s).

\subsection{QPO Classification}
\label{sec:qpo}

QPOs can be characterized by three main properties that can be measured from the power spectra: the frequency of the QPO, the rms variability and the quality factor Q (Q=frequency/FWHM). LFQPOs from black holes are classified into one of three types: A, B or C \citep{Casella2005}.\footnote{These phenomenological classifications are used because of the lack of a clear understanding of the physical origins of these QPOs.} In this paper, we will also look at how the phase coupling properties of the QPOs can be used to help understand them.

Type A QPOs generally occur in the frequency range of $\sim$6-8 Hz. These QPOs are typically broad (Q<3), show low rms variability and do not have any observed harmonic structures. Type B QPOs occur in the frequency range of $\sim$0.5-6.5 Hz, with Q>6. These sometimes show weak subharmonic or harmonic structures. Both Type A and B QPOs occur when the source is in the SIMS. 

Type C QPOs appear at frequencies between $\sim$0.1 and $\sim$10 Hz, and are usually narrow (Q $\sim$ 7-12) with high rms variability. This class of QPOs are often observed along with the presence of higher harmonics, with a weak sub-harmonic feature sometimes being present. These QPOs are seen when the source is in LHS or HIMS. For a more detailed overview of the properties of the type A,B and C QPOs seen in GX~339-4, we refer the reader to \cite{Motta2011}.  

\begin{table*}
\caption{Details of the RXTE observations used in this paper. The full table is available online.}
\centering
	\begin{tabular}{|l c c c c c c c c c|}
    \hline
	No. & ObsId & MJD & Outburst & Hardness & QPO frequency & Type & State & Mean & Pattern \\ 
		&	& & 
        & Ratio$^a$	& [Hz]$^a$ &	& 	& Count& \\ 
    \hline
1 & 40031-03-02-05 & 52388.054 & 2002 & 0.766 & 0.20 & C & LHS & 1055 & Web* \\  
2 & 70109-01-05-01G & 52391.318 & 2002 & 0.763 & 0.22 & C & HIMS & 1055 & Web* \\ 
3 & 70109-01-06-00 & 52400.83 & 2002 & 0.697 & 1.26 & C & HIMS & 1120 & Web* \\  
4 & 70108-03-01-00 & 52400.853 & 2002 & 0.694 & 1.30 & C & HIMS & 1111 & Web \\  
5 & 70110-01-10-00 & 52402.492 & 2002 & 0.562 & 4.20 & C & HIMS & 1091 & Web \\  
    ... \\
    \hline
        \multicolumn{3}{l}{$^a$ Values taken from \citep{Motta2011}}\\
    \multicolumn{3}{r}{$^*$ QPOs with $f <$2 Hz classed as web if pattern is unclear}\\
	\end{tabular}
\label{table:obs}
\end{table*}

\section{Results}
\label{sec:results}

\subsection{Bicoherence Patterns}

For classifying the different patterns seen in the bicoherence, we follow the convention used in \cite{Maccarone2011}. Below, we include a brief description of the different patterns observed in GX~339-4. The `cross' pattern has not been observed from this source. We also analyse and describe the bicoherence of type B QPOs. 

\subsection{Type C QPOs}
\begin{figure}
	\includegraphics[width=\columnwidth]{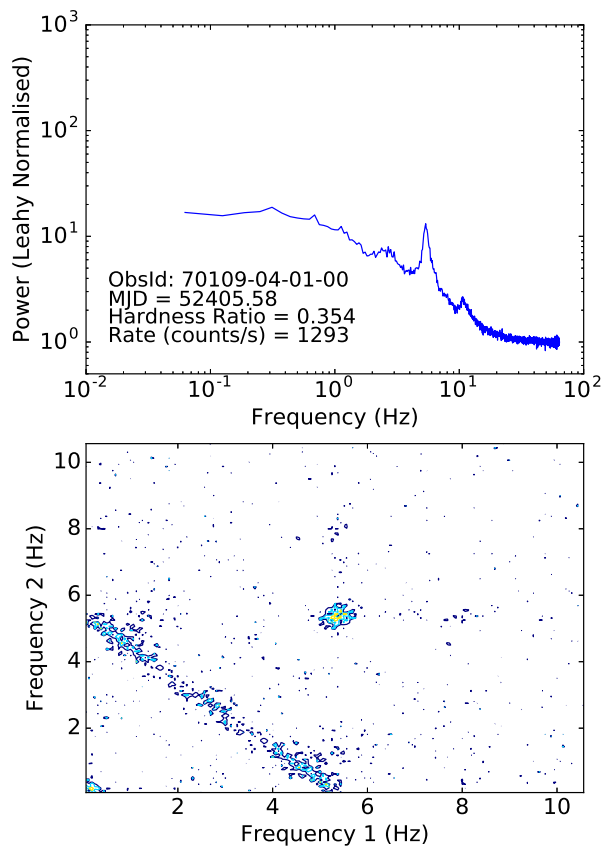}
    \caption{The (a) Power spectrum and (b) the bicoherence plot showing the 'hypotenuse' pattern from the observation 70109-04-01-00. The colour scheme of log$b^2$ is as follows: dark blue:-2.0, light blue:-1.75, yellow:-1.50, red:-1.25}
    \label{fig:hypotenuse}
\end{figure}

\subsubsection{The `hypotenuse' pattern}

When the `hypotenuse' pattern is observed (see Fig~\ref{fig:hypotenuse}) , a high bicoherence is seen when the two noise frequencies $f_1$ and $f_2$ add up to the frequency of the QPO, making a diagonal line. This pattern also shows a region of high bicoherence when $f_1 = f_2 = f_{QPO}$. This demonstrates that there is power at the frequency of the QPO and the first harmonic and that there is coupling between these two frequency components. Essentially, if there is power in the $f_1=f_2$ part of the diagram, this indicates that significant power comes in the form of a non-sinusoidal waveform (see Section~\ref{sec:waveform} for details on the QPO waveform).  If harmonics appeared in the power spectrum for frequencies $f_1$, $f_2$ and$f_1+f_2$, but stringent upper limits existed on the power of the bispectrum, that would argue that the harmonics represented different normal modes of a system which were being excited independently (consider, e.g. the case of a guitar string being plucked exactly at several of its nodes with totally uncorrelated driving forces).

\subsubsection{The `web' pattern}

\begin{figure}
	\includegraphics[width=\columnwidth]{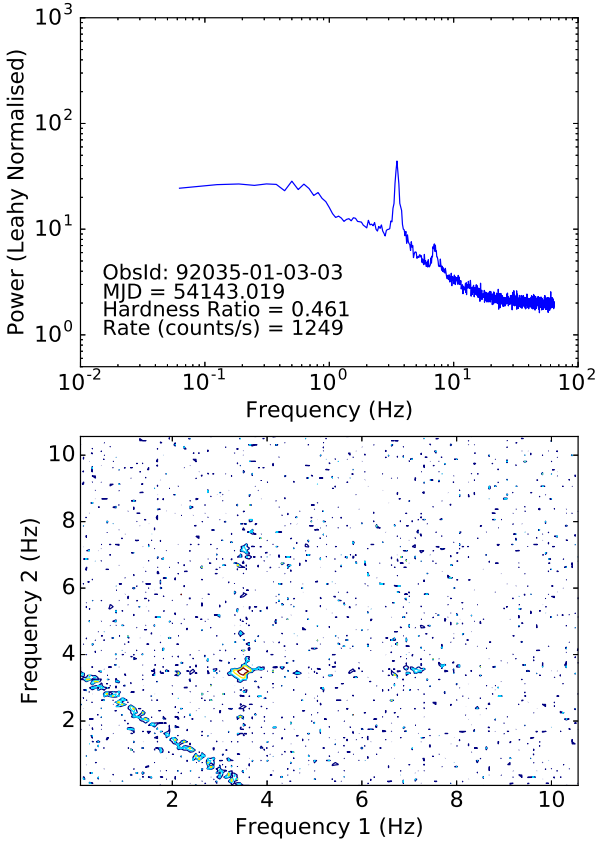}
    \caption{The (a) Power spectrum and (b) the bicoherence plot showing the 'web' pattern from the observation 92035-01-03-03. The colour scheme of log$b^2$ is as follows: dark blue:-2.0, light blue:-1.75, yellow:-1.50, red:-1.25}
    \label{fig:web}
\end{figure}

The `web' pattern (see Fig~\ref{fig:web}) shows features seen in both the hypotenuse and the cross pattern. The diagonal line of high bicoherence when $f_1 + f_2 = f_{QPO}$ is present. Also present is a vertical (and horizontal) streak where $f_1$ (or $f_2$) corresponds to $f_{QPO}$ and the other frequency is noise. The strength of bicoherence falls off at frequencies above 2$f_{QPO}$. Regions of high bicoherence can also be seen when the fundamental frequency interacts with the second harmonic to produce power at higher harmonics.

\subsection{Type B QPOs}

\begin{figure}
	\includegraphics[width=\columnwidth]{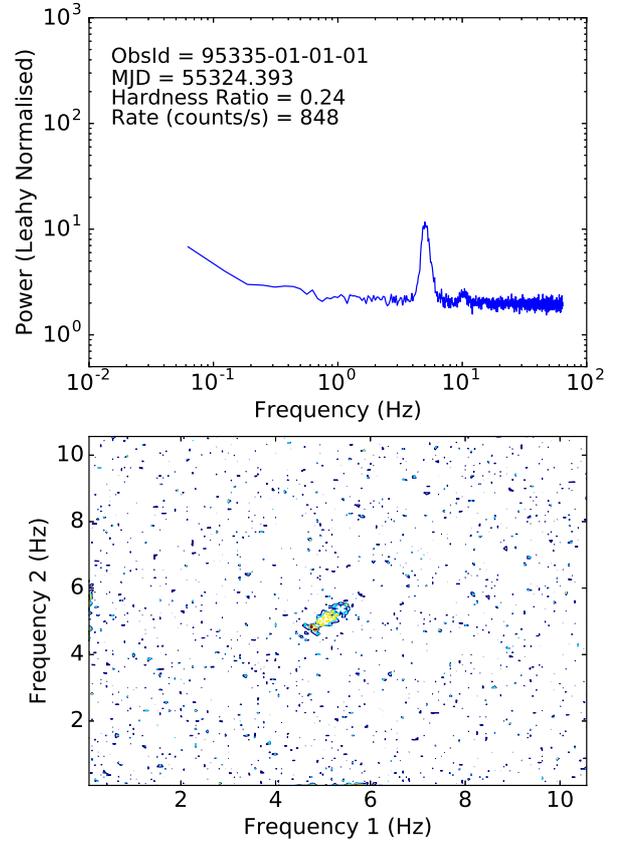}
    \caption{The (a) Power spectrum and (b) the bicoherence plot for a type B QPO from the observation 95335-01-01-01. The colour scheme of log$b^2$ is as follows: dark blue:-2.0, light blue:-1.75, yellow:-1.50, red:-1.25}
    \label{fig:blob}
\end{figure}

When a type B QPO is present, (see Fig~\ref{fig:blob}) a high bicoherence is seen where both $f_1$ and $f_2$ are equal to $f_{QPO}$, indicating the presence of a second harmonic (often visible in the power spectrum). However, it does not show any coupling with the broad band noise frequencies. When a sub-harmonic is present in the power spectrum, its presence is also indicated by a feature in the bicoherence plot where $f_1$ and $f_2$ are equal to half of $f_{QPO}$. The diagonal elongation of the region from the bottom left to the top right is most likely a result of the frequency drift of the QPO over the length of the observation. This appears to be the only pattern (other than pure harmonic correlation) seen with a type B QPO. 

\subsection{Type A QPOs}

\begin{figure}
	\includegraphics[width=\columnwidth]{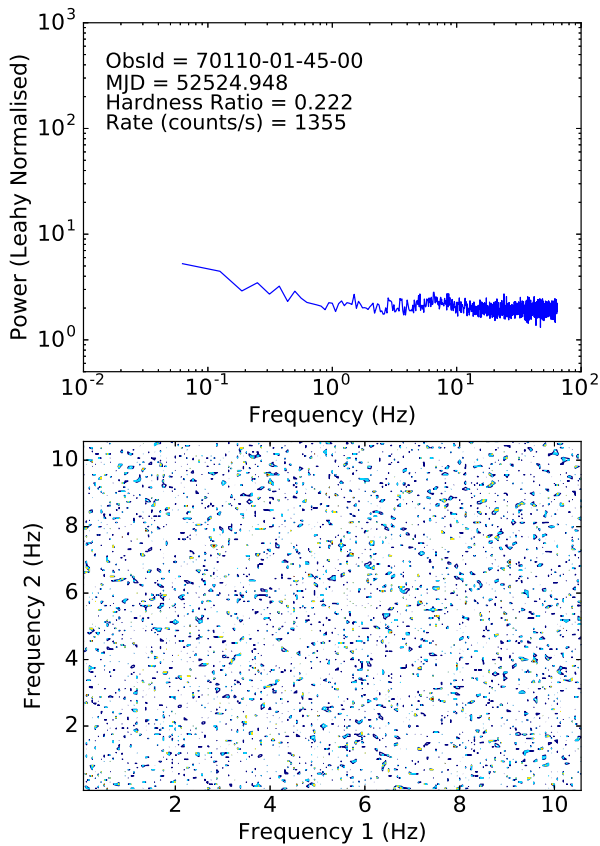}
    \caption{The (a) Power spectrum and (b) the bicoherence plot for a type A QPO from the observation 70110-01-45-00. The colour scheme of log$b^2$ is as follows: dark blue:-2.0, light blue:-1.75, yellow:-1.50, red:-1.25. No evidence for phase coupling between the QPO and a harmonic is seen.}
    \label{fig:typea}
\end{figure}

The bicoherence plots of type A QPOs do not show any discernible patterns such as in the case of type B or C QPOs. To illustrate this with a quantitative example, we use the observation 70110-01-45-00 which has a QPO at 7.2Hz (see Fig.~\ref{fig:typea}), and examine the mean value of the bicoherence in different regions of the plot using a frequency resolution of 1/16Hz. Along the diagonal region where $f_1$+$f_2$ are in the range 113 to 117 times the frequency resolution, the bicoherence has a mean value of 0.029$\pm$0.129. In the region where both $f_1$ and $f_2$ are in the range 113 to 117 times the frequency resolution, the bicoherence has a mean value of 0.012$\pm$0.011. The value of the bias in the bicoherence for this observation is 0.011. Type A QPOs are not accompanied by either subharmonic or higher harmonic features. 

\subsection{Evolution during state transition}

During the course of an outburst,the source often moves from a hard state, which is often accompanied by a type C QPO to a soft state, which often shows the presence of a type A or type B QPO. In this paper, we present plots of observations during the 2007 outburst. The details of all the observations analysed, along with their respective bicoherence plots, are available in the online version of this paper. As the source softens
and the QPO evolves to higher frequencies, the bicoherence pattern changes, transitioning smoothly from the `web' pattern, to the `hypotenuse'. As the source enters the SIMS and a type B QPO is seen, the bicoherence abruptly changes and no longer shows coupling between the QPO and the broadband noise.

\begin{figure*}
  \centering
  \subfloat{\includegraphics[width=0.75\columnwidth]{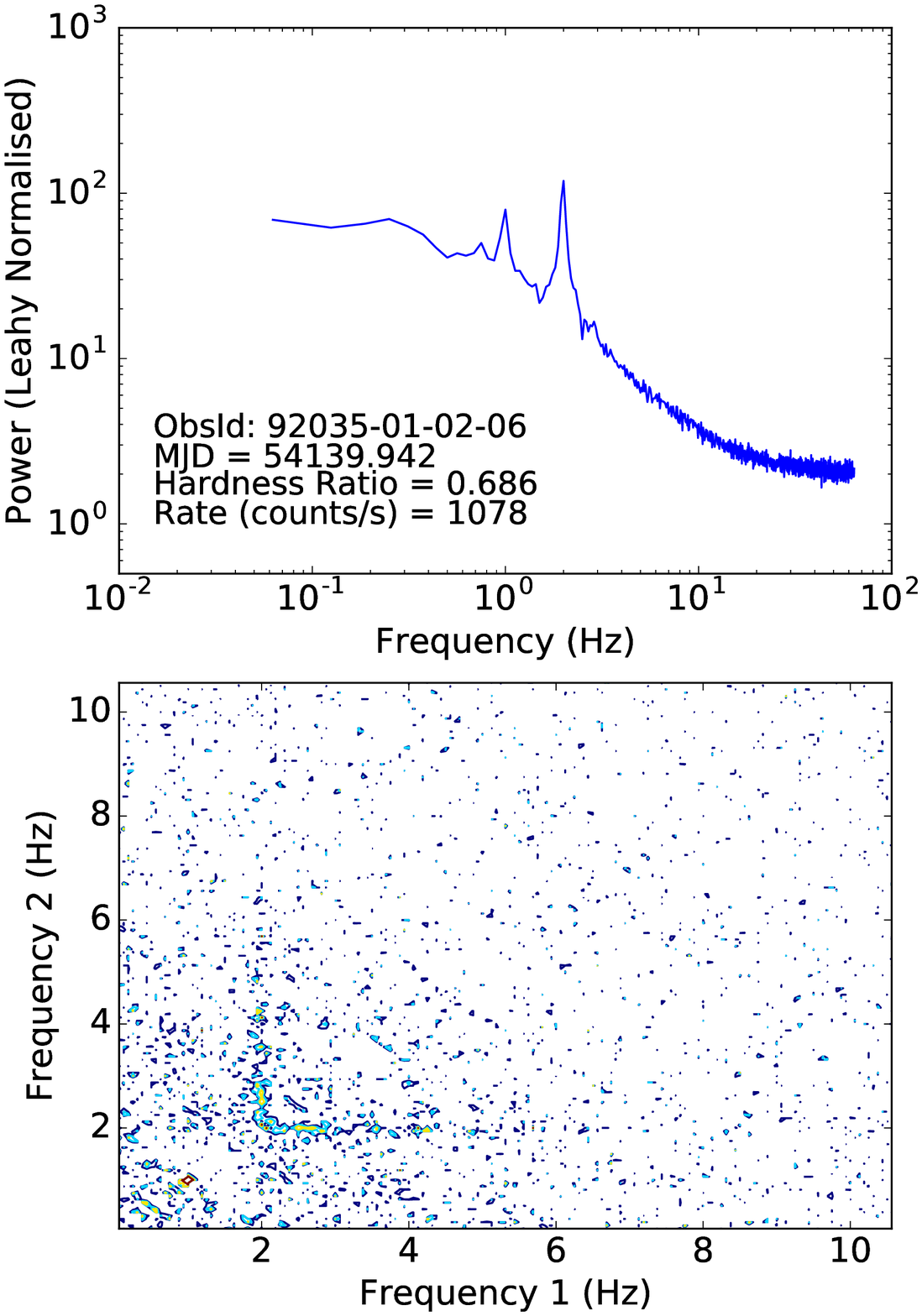}}
  \subfloat{\includegraphics[width=0.75\columnwidth]{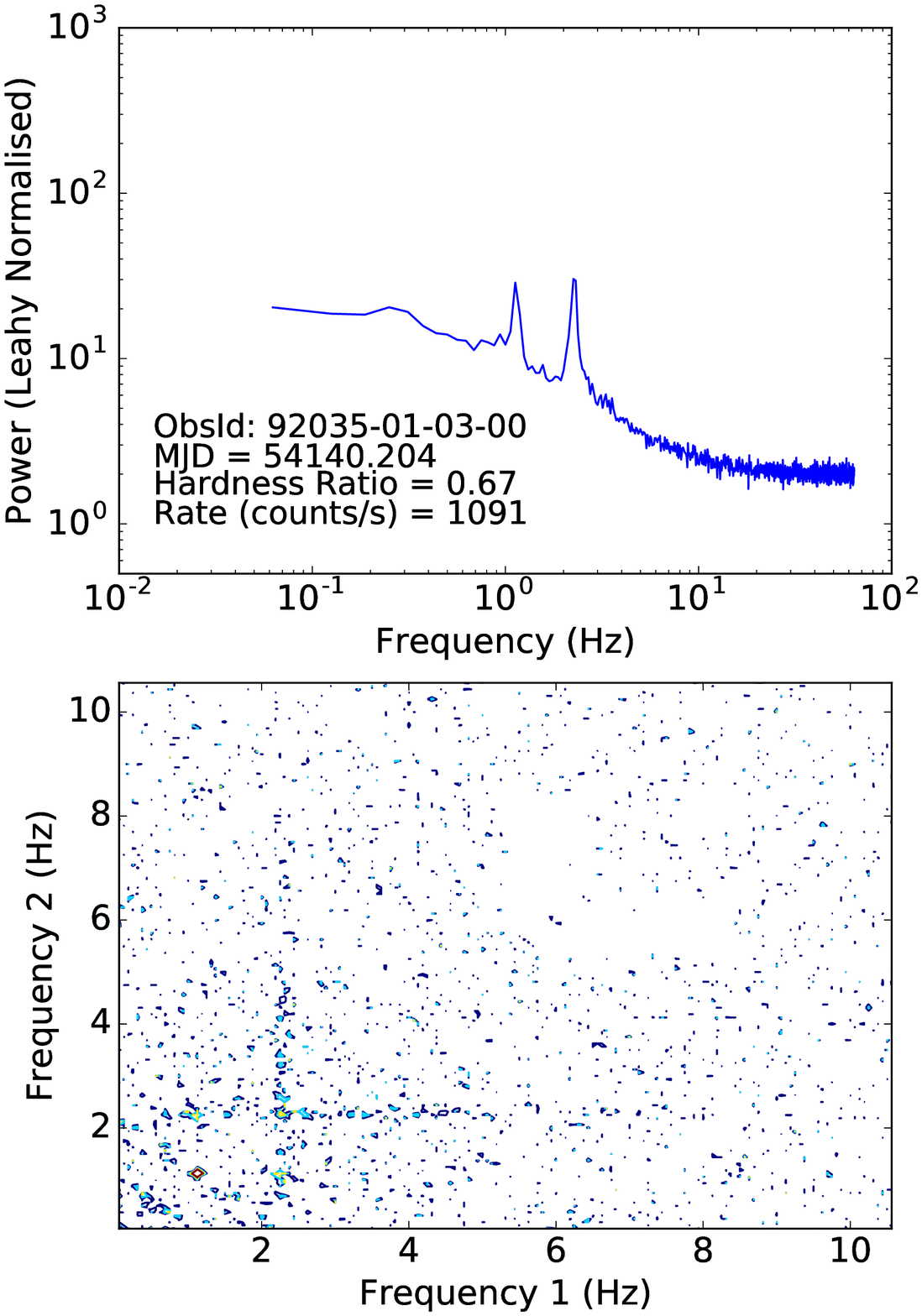}}
  \subfloat{\includegraphics[width=0.75\columnwidth]{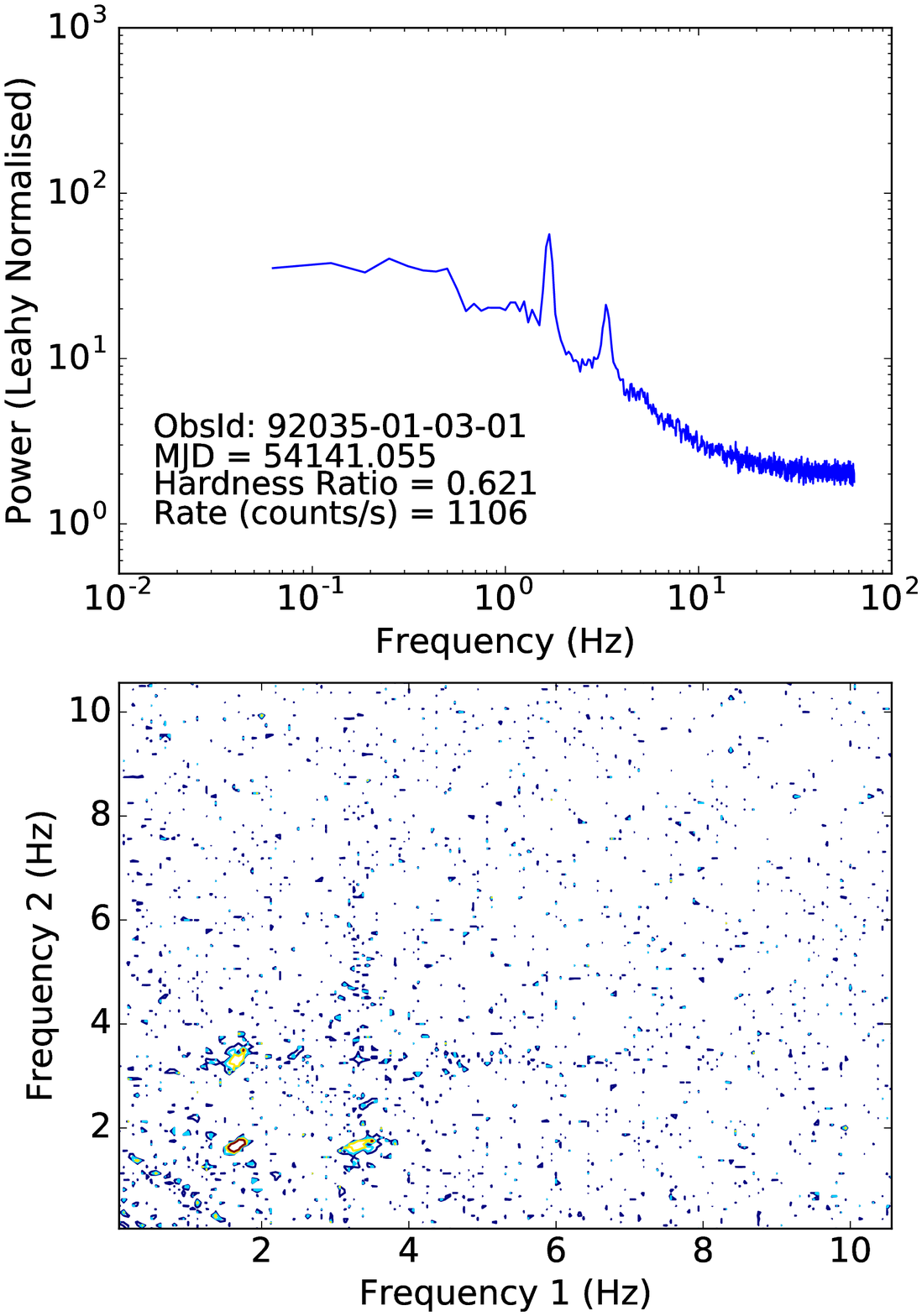}}\\
  \subfloat{\includegraphics[width=0.75\columnwidth]{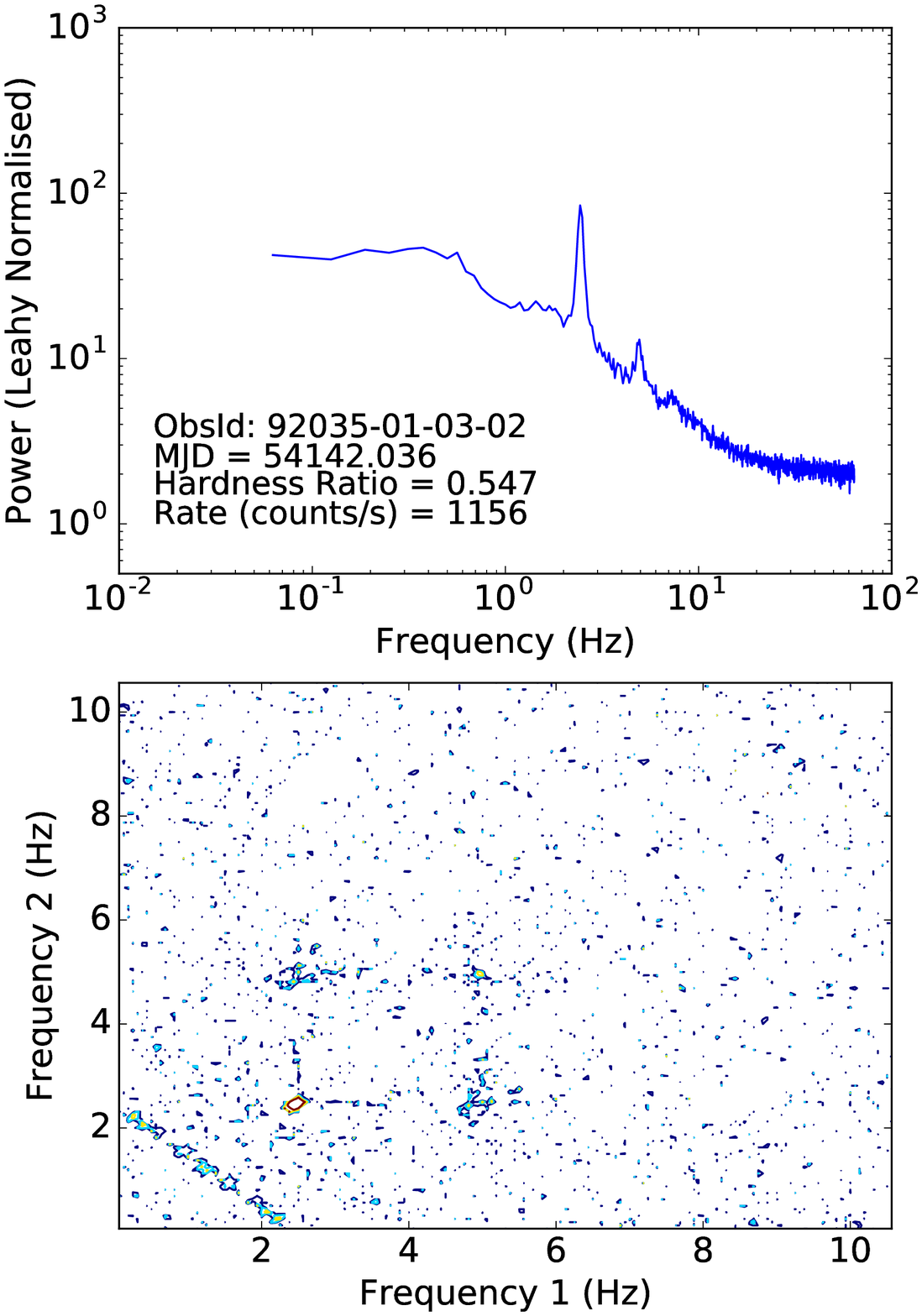}}
  \subfloat{\includegraphics[width=0.75\columnwidth]{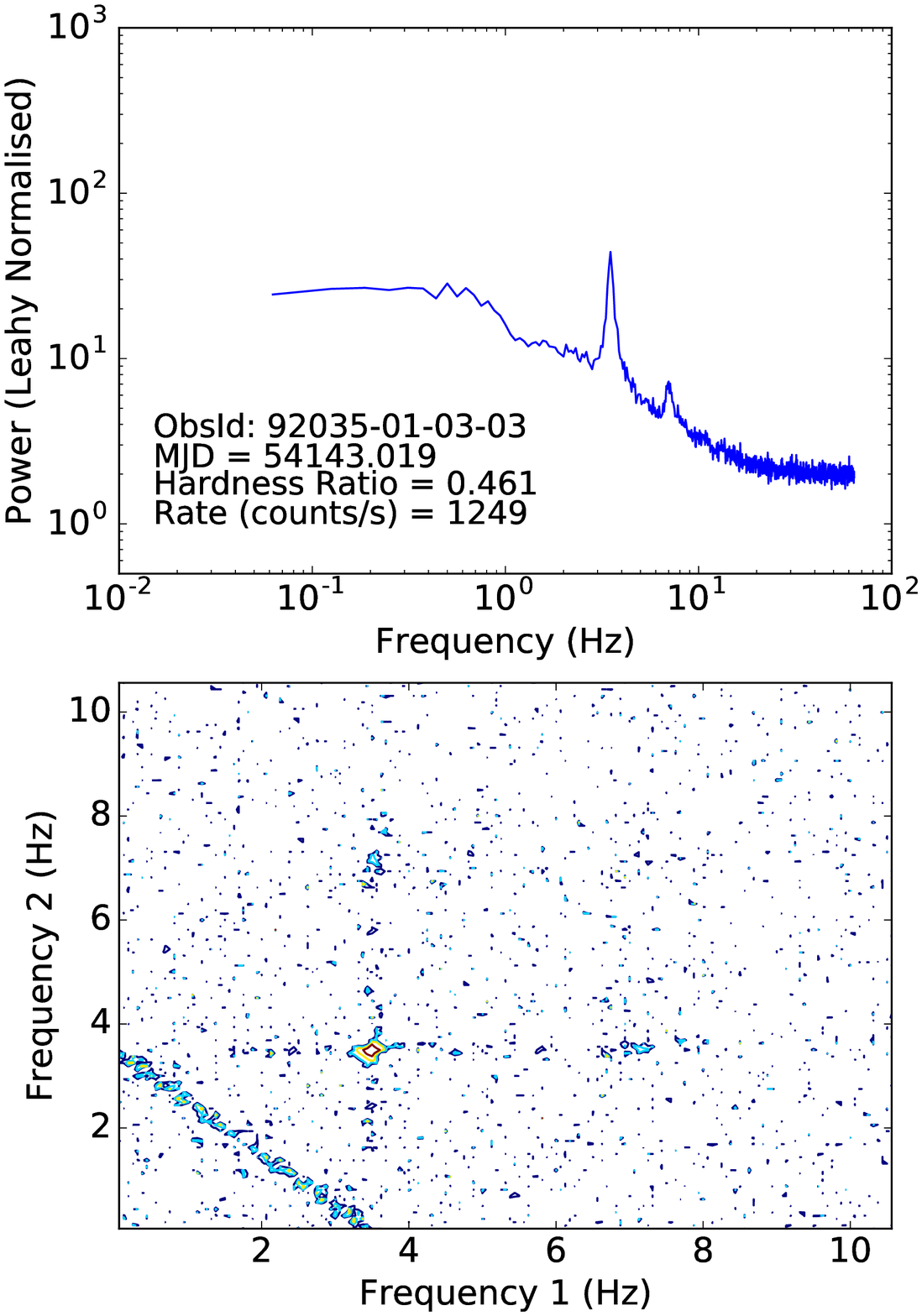}}
  \subfloat{\includegraphics[width=0.75\columnwidth]{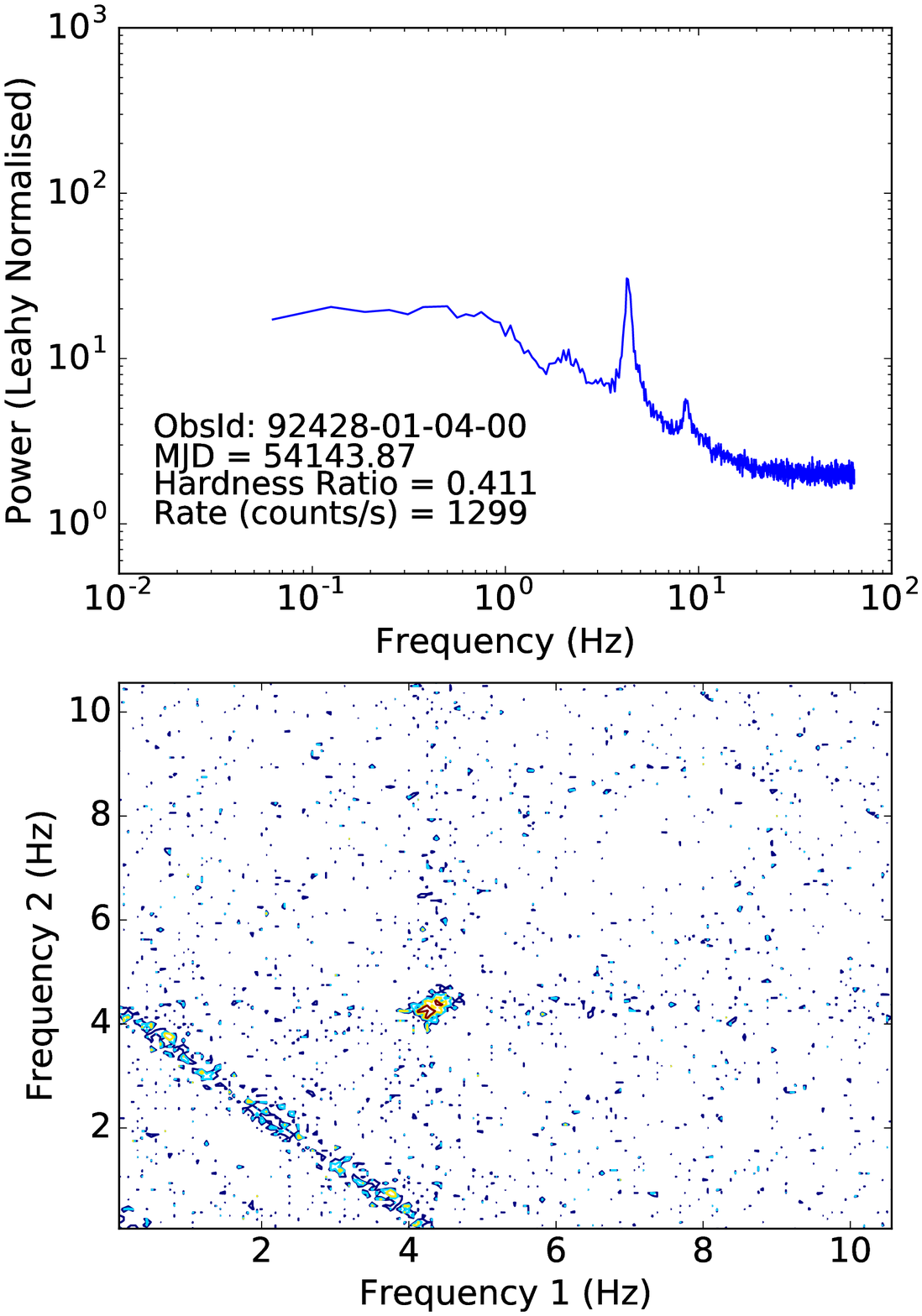}}
  \caption{The power spectrum and the bicoherence plot for multiple observations of GX~339-4 during the 2007 outburst. The colour scheme of log$b^2$ is as follows: dark blue:-2.0, light blue:-1.75, yellow:-1.50, red:-1.25. As the source evolves from HIMS to SIMS, the bicoherence pattern gradually evolves from a `web' to a `hypotenuse' state. The bicoherence plots of all the observations listed in Table 1 can be found in the online version of this paper. }
  \label{fig:evolution}
\end{figure*}

The strength of the bicoherence in the vertical/horizontal streaks steadily falls off as the power in the diagonal structure increases until the `hypotenuse' pattern is seen. This transition occurs 3-6 days before the source enters the SIMS. This change in the bicoherence pattern is observed most clearly during the 2007 outburst (see Fig.~\ref{fig:evolution}), but is also seen during the 2002 and 2010 outbursts. Due to a combination of the observations having short exposure and/or low count rate, it was not possible to see if this change occurs during the 2004 outburst.

\begin{figure}
	\includegraphics[width=\columnwidth]{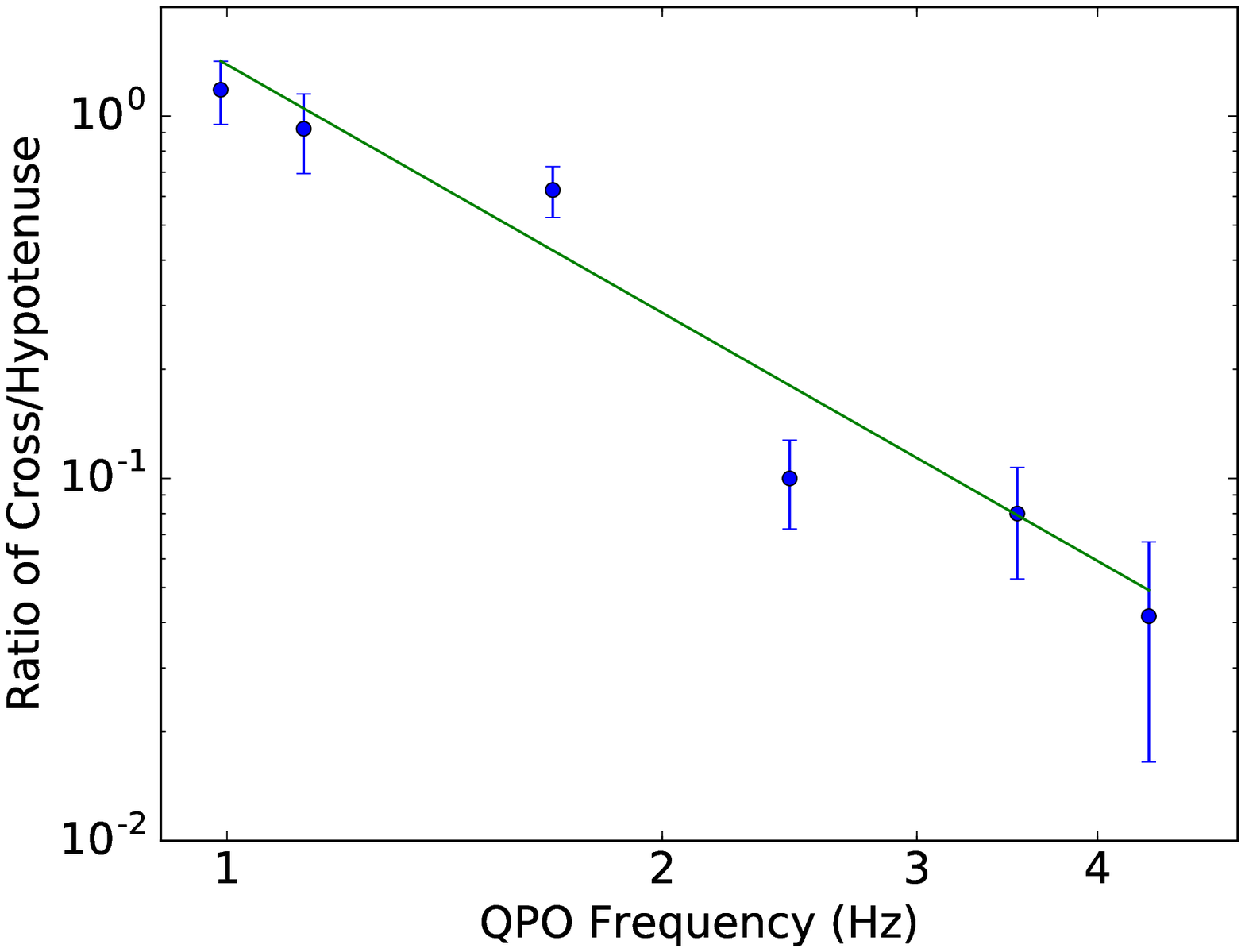}
    \caption{The ratio between the mean value of the bicoherence along the 'cross' and the `hypotenuse'. It can be seen that the strength of the bicoherence along the `cross' with respect to that along the `hypotenuse' decreases during the transition. The solid line shows a linear fit with a reduced $\chi^2$ of 1.9, and a null hypothesis of 0.11 indicating that a simple power law is an adequate description of the data.}
    \label{fig:ratio}
\end{figure}

Figure~\ref{fig:ratio} shows the relative strength of the bicoherence along the horizontal and vertical `cross' pattern and the diagonal `hypotenuse' pattern as a function of QPO frequency. The ratio was calculated using the mean value of the bicoherence along the cross divided by the mean value of the bicoherence along the hypotenuse, and this was done for each of the 6 observations shown in Figure~\ref{fig:evolution}. While estimating the bicoherence along the `cross', the regions where $f_1$ or $f_2$ were equal to $f_{QPO}$ or $2f_{QPO}$ were excluded, as these regions would be dominated by the coupling between the QPO and the harmonic. Additionally, for estimates of the bicoherence along the `hypotenuse', values from the lowest frequency bins were excluded. Due to the length of the observations, the source was not observed for many cycles at these low frequencies, and thus the bicoherence values in these bins are not reliable. A clear downward trend is seen in Figure~\ref{fig:ratio}, indicating that the strength of the bicoherence along hypotenuse increases relative to that along the cross.

The possibility of quasi periodic oscillations being produced by a damped, forced oscillator has been previously suggested \citep{Maccarone2011}. It is also well known that a damped, forced, non-linear oscillator, also known as a Duffing Oscillator \citep{Duffing1918}, is able to produce quasi-periodic oscillations. [See e.g \cite{Phillipson2018} for an application of the Duffing oscillator to longer timescale variability in an X-ray binary.] We find that an increase in the driving frequency of the oscillator, results in the bicoherence pattern of the resulting lightcurves evolving in a manner similar to that shown in Fig.~\ref{fig:evolution}. However, the connection between the parameters of the oscillator and physical quantities (such as viscosity, optical depth and mass transfer rate) is presently unclear.

It is also worth noting that the amplitude type B and C QPOs show dependence on the inclination angle of the source \citep{Motta2015}, with the type C QPOs being stronger in the high inclination (edge on) systems, and type B QPOs being stronger in the low inclination (face on) systems. In future work, we will present a comprehensive study of the effects of inclination on the bicoherence. 

\subsection{Reconstructing the QPO waveforms}
\label{sec:waveform}

It has previously been shown by \citet{Ingram2015} that the phase difference between the first two QPO harmonics varies around a well defined average, and that this can be used to reconstruct the QPO waveform.

The biphase holds information about the shape of the underlying waveform such as the skewness of the flux distribution and the asymmetry of the time series \citep{Maccarone2013}. It must be noted that a value for the biphase always exists, even when no phase coupling is present. Thus the biphase only contains useful information in regions of statistically significant bicoherence. The bispectrum provides a formal approach with well understood statistical properties to addressing the question of phase difference between harmonics. The value of the biphase in the region $f_1 = f_2 = f_{QPO}$ gives the phase difference between the fundamental and second harmonics. This can be used in a similar fashion to reconstruct the QPO waveform. Since the biphase contains information about the phase difference between any set of three frequencies, a waveform can also be reconstructed with any phase information from any higher harmonics that are present.

\begin{figure*}
  \centering
  \subfloat{\includegraphics[width=0.75\columnwidth]{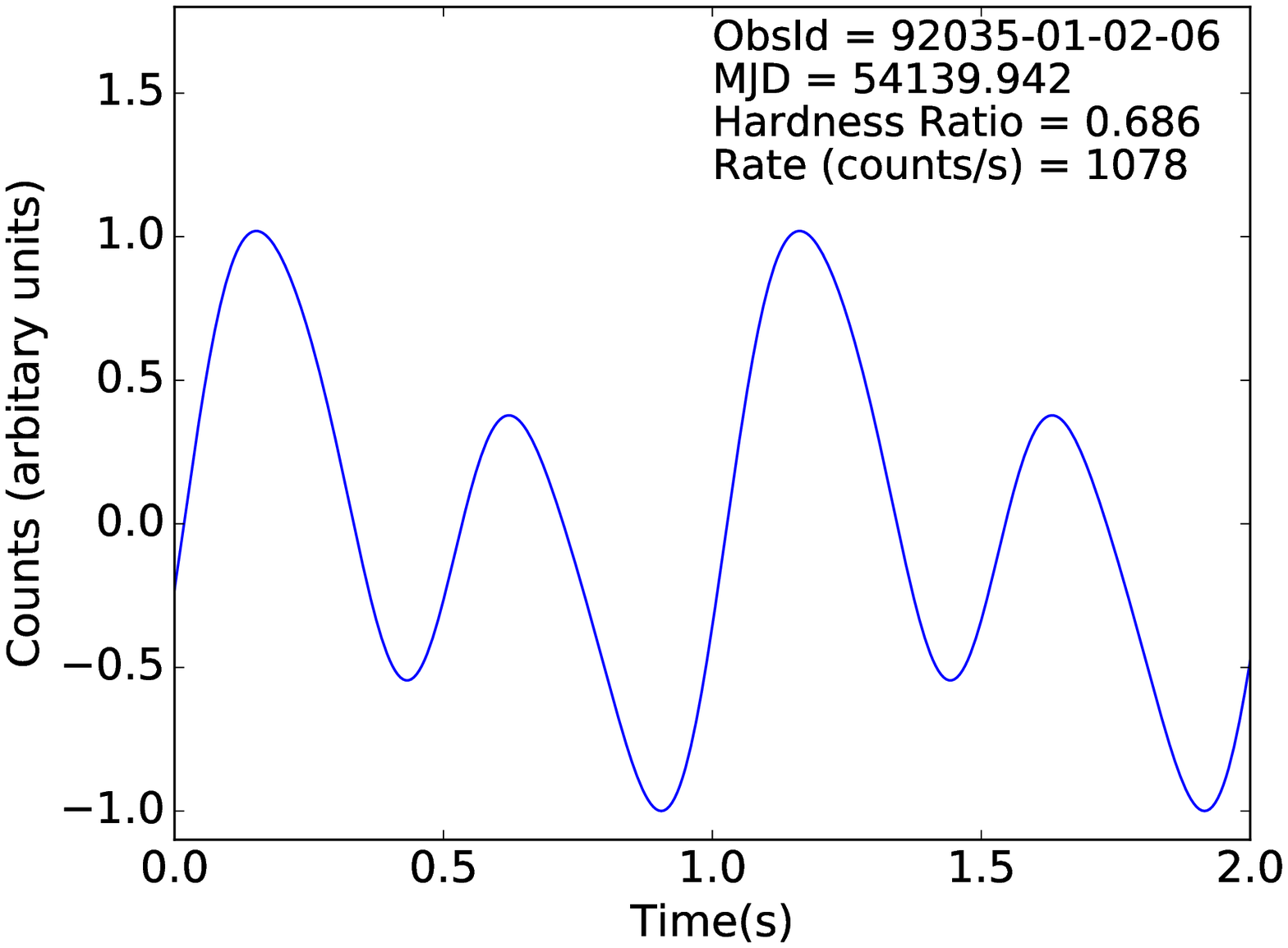}}
  \subfloat{\includegraphics[width=0.75\columnwidth]{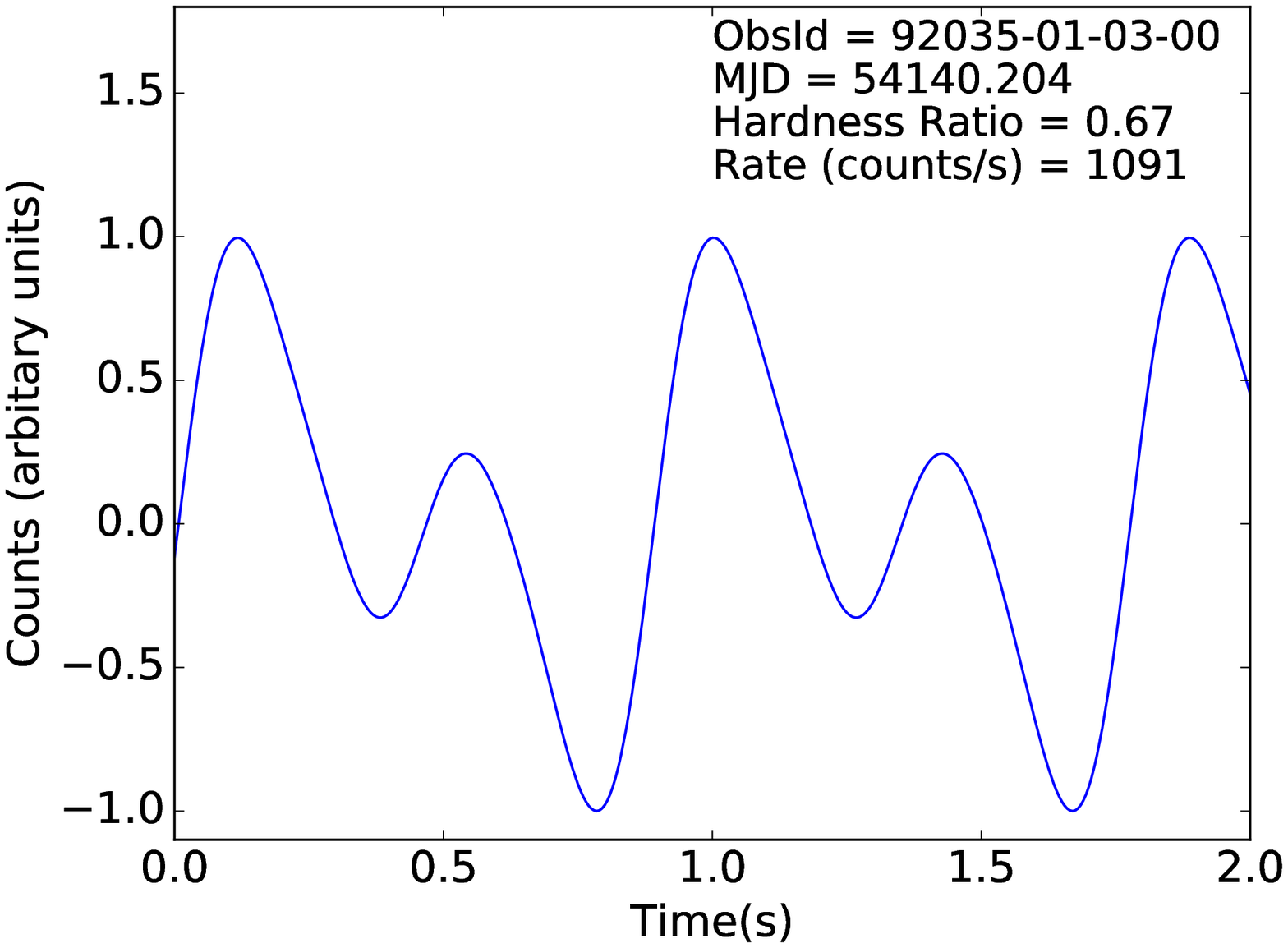}}
  \subfloat{\includegraphics[width=0.75\columnwidth]{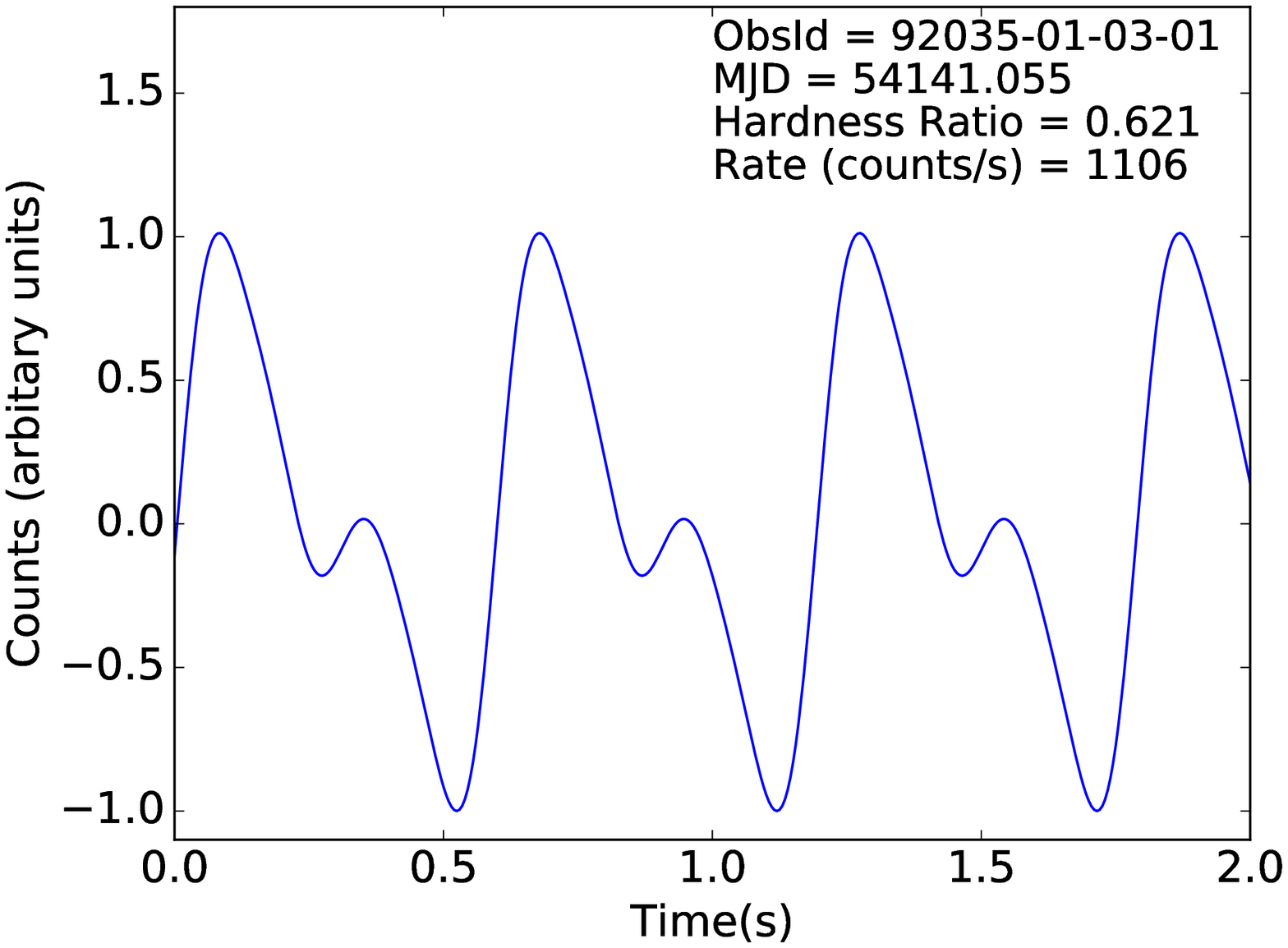}}\\
  \subfloat{\includegraphics[width=0.75\columnwidth]{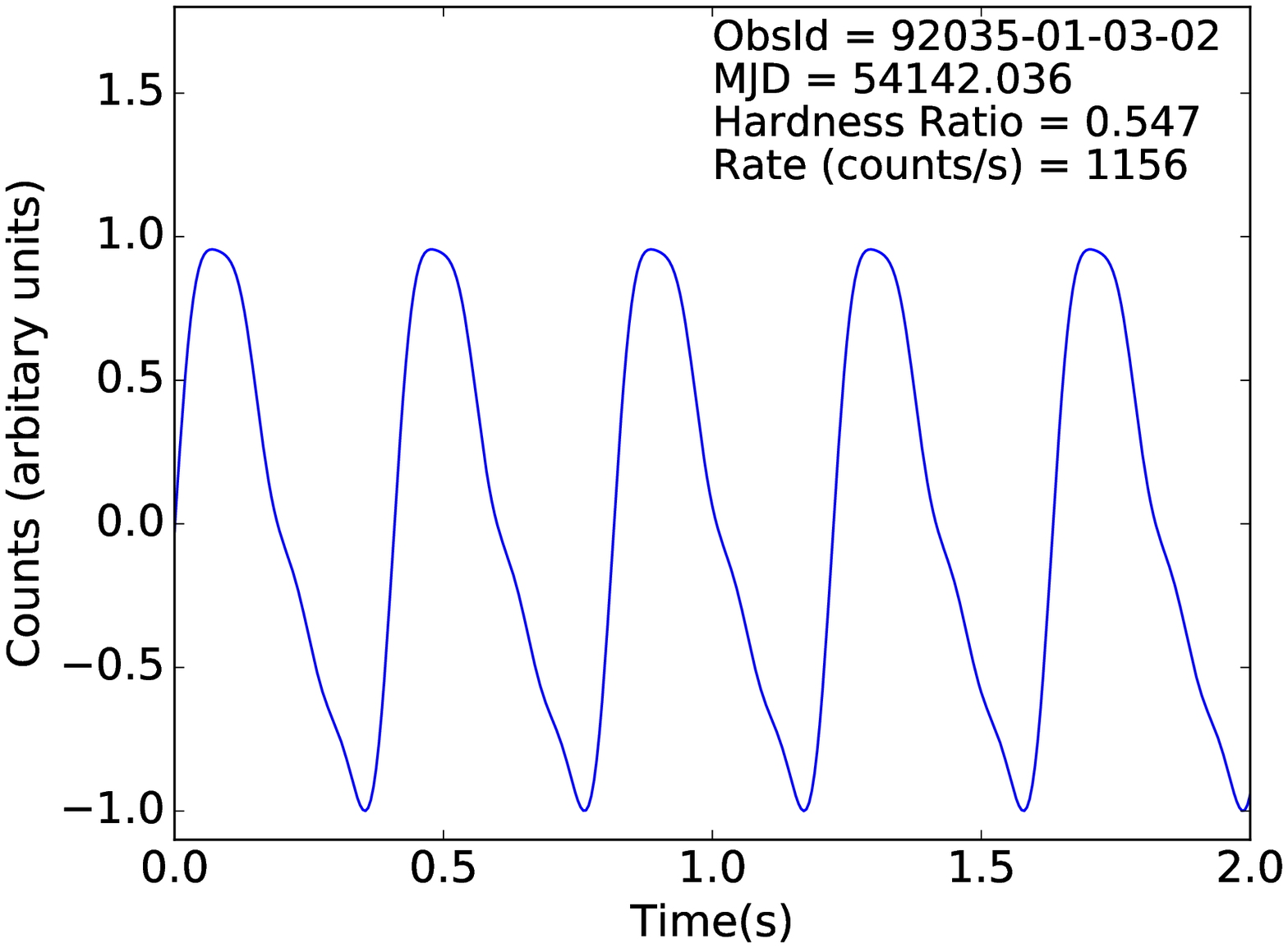}}
  \subfloat{\includegraphics[width=0.75\columnwidth]{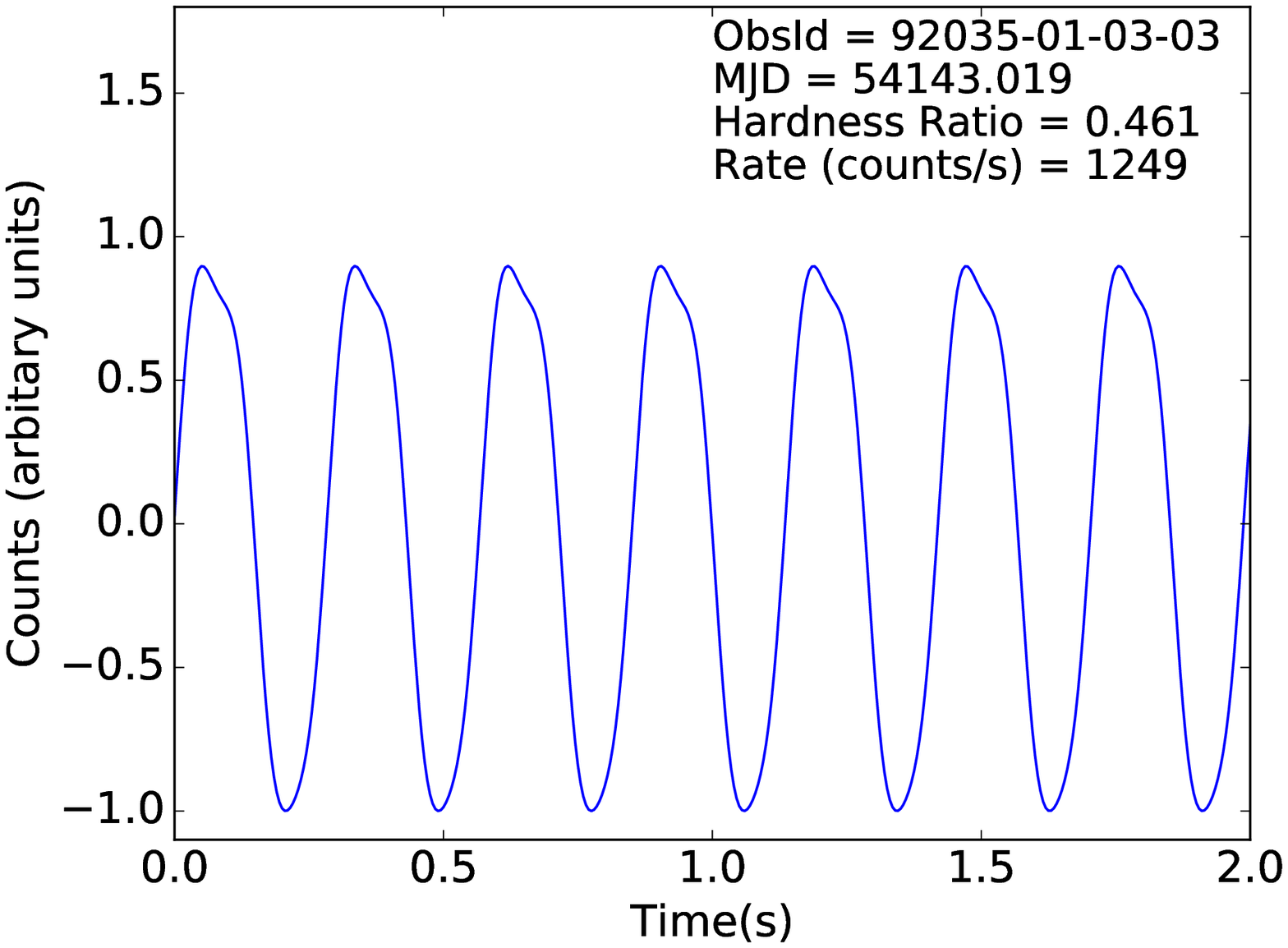}}
  \subfloat{\includegraphics[width=0.75\columnwidth]{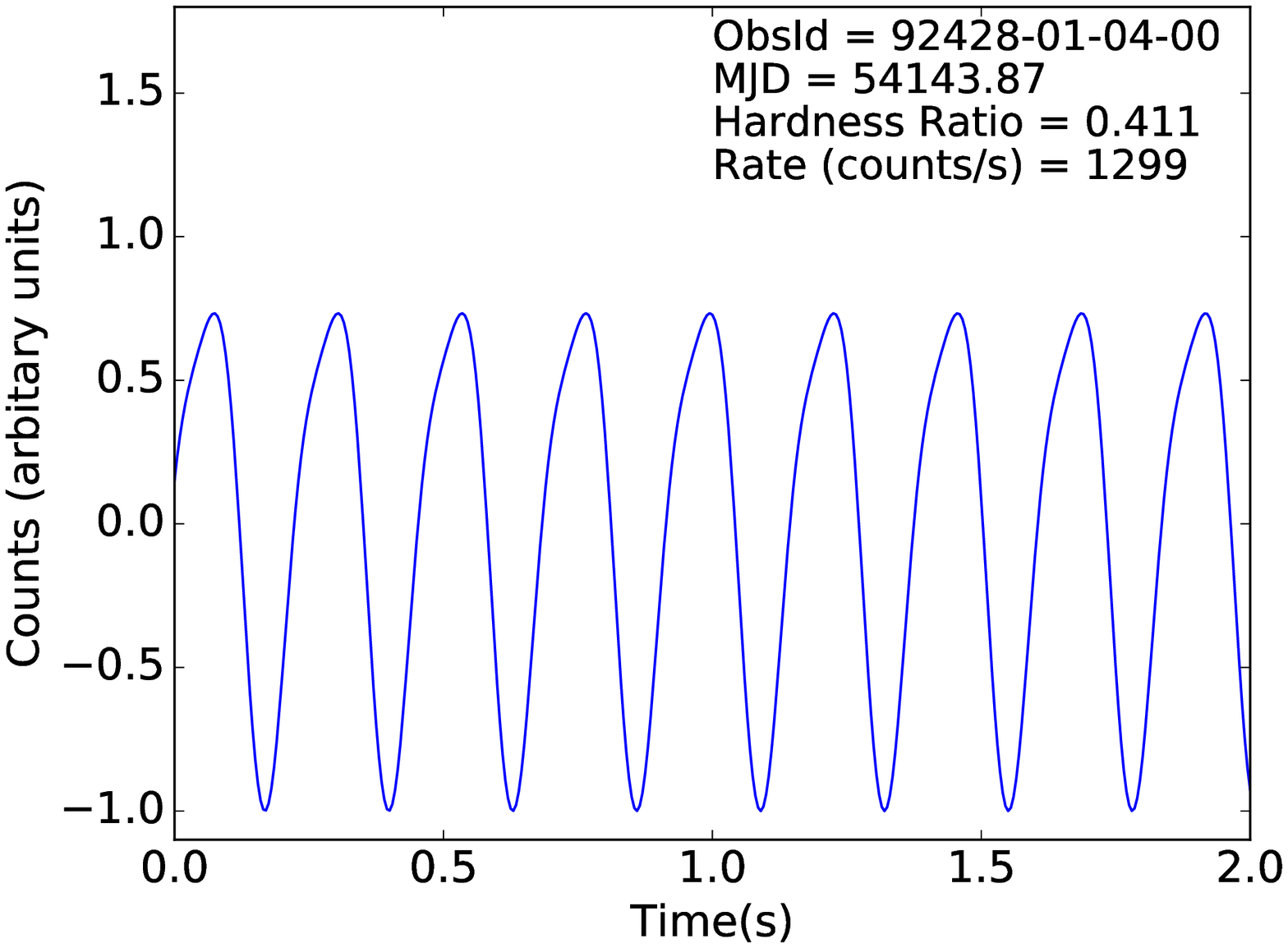}}
  \caption{The waveforms of QPOs from the observations in Fig~\ref{fig:evolution}, reconstructed using the values of the biphase. }
  \label{fig:waveform}
\end{figure*}

Fig~\ref{fig:waveform} shows the waveforms of QPOs from the 6 observations shown in Fig~\ref{fig:evolution}. The relative amplitude of the harmonics was obtained from the height of the peaks in the power spectrum, while the relative phase difference between the harmonics was obtained from the biphase. It can be seen that the phase difference between the harmonics evolves between different observations, leading to a change in the QPO waveform, with the waveform changing from two maxima per cycle to one maxima per cycle.

\section{Discussion}
\label{sec:discussion}

\subsection{Physical Interpretation}
\label{sec:physical}

In regions where there is a large bicoherence, phase coupling is present among the 3 frequencies involved. The gradual change in the pattern of the bicoherence over the state transition indicates a gradual change in the accretion disk.

Here, we suggest that this behaviour could be explained by an increase in the optical depth of the corona as the sources transitions from the HIMS to the SIMS. In this scenario, during the HIMS the disk component is low and with an optically thin ($\tau \lesssim $ 1) corona. The QPO, caused by the precession of the inner accretion flow (\cite{Stella1998}, \cite{Ingram2011}) , modulates the high frequency variability originating in the inner regions of the accretion disk. This modulation produces the vertical and horizontal lines in the bicoherence. Additionally, the inverse Compton scattering of the soft photons from the thin disk by the corona, causes the low frequency variability of the disk to modulate the QPO. This amplitude modulation causes the weak diagonal line that is seen in the bicoherence. Overall, this produces the web pattern that is seen in the HIMS.

As the source moves towards the SIMS, the optical depth of the corona increases. In this state, the disk blackbody component dominates, with a larger number of soft seed photons being upscattered. This leads to a stronger modulation of the QPO from the low frequency variability, giving rise to the prominent hypotenuse pattern. If the diffusion time of photons through the optically thick corona is larger than the time scales of the high frequency variability, the HF variability is smeared out. This weakens the previously observed vertical and horizontal lines in the bicoherence. Overall this leads to the hypotenuse pattern that is observed in the SIMS.

It has been shown by \cite{Kylafis1987} that there is significant smearing of  oscillations of frequency $f$ by electron scattering when in a spherical cloud of radius R with an optical depth of $\tau$ if $2\pi f \tau R/c \gg 1$. Assuming a 10$M_{\odot}$ black hole with a comptonising region of R=50$R_s$ \footnote{Schwarzschild Radius $R_s= 2GM/c^2$}, an optical depth of $\tau \gg 3.2$ is sufficient to smear out any variations on timescales faster than f$\sim$10Hz. 

A similar thermal Comptonisation model that invokes an increase in the optical depth of the Comptonising region was proposed by \citet{Nobili2000}. This model was used to reproduce the inversion of the sign of the phase lags as the spectrum softens and the QPO frequency increases above 2Hz as seen in GRS 1915+105 \citep{Reig2000}. Sign reversal of the phase lag has also been seen in XTE J1550-564 \citep{Remillard2002}. Detailed studies of the phase lags of QPOs from XTE J1550-564 have also revealed more complex behaviour such as a difference in the sign of the phase lags between the QPO fundamental and the harmonic \citep{Cui2000}. Additionally, the phase lag behaviour of Type C QPOs also show an inclination dependence \citep{VandenEijnden2017}. Thus a comprehensive study of the evolution of the phase lags in this model would require the computation of a detailed ray tracing model.

In principle, one can assess whether the changes in the optical depth involved here are consistent with model fits of the spectral energy distribution. The changes in the energy spectrum depend on the interplay between the temperature of the electrons in the corona and the optical depth. A recent analysis of the spectra of GX 339-4 in the hard state over multiple outbursts \citep{Garcia2015} found that the temperature of the corona decreases with increasing luminosity, while the optical depth increases for a nominal value of the photon index.  Similarly, a study of the 2007 outburst of GX 339-4 by \citet{Motta2009} showed a monotonic decrease of the high energy cut-off during the brightening of the hard state, while the photon index stayed roughly constant. However, during the intermediate state, the cut off energy and the photon index increased, implying a decrease in the optical depth if a purely thermal electron distribution is assumed. But given the clear trends in the hard state, it is reasonably likely that this is the emergence of a non-thermal electron distribution in this state imitating an increase in the cut off energy. Indications of the presence of such a non-thermal component have been seen in Cygnus X-1 in its soft state \citep{McConnell2002}. Because often spectral cutoffs are inferred without high signal to noise data extending well beyond the cutoff energy, it is often hard to estimate the effects of non-thermal components to the electron distribution on the spectrum. In fact, it may be the case that after a full understanding is developed, timing-based approaches like what we start to develop in this paper may become more robust estimators for the geometrically thick hot region's optical depth than spectra.

With this in mind, we used the RXTE standard products data to fit Observation 23 from \citet{Motta2009}, which is the intermediate state observation that is most secure in having a higher cutoff energy than the last few hard state observations before the intermediate state began.  We used the same baseline model as \citet{Motta2009}, but then added a power law component to account for our putative non-thermal Comptonisation component.  When doing this, we find a spectral index of $\Gamma=2.46$ for this extra power law component, with the non-thermal component  dominating the total flux across the bandpass.  If we freeze the index to 2.2, then the non-thermal component dominates only above 90 keV.  In the former case, the cutoff energy for the component with a cutoff drops to 17 keV, while in the latter case it drops to 50 keV.  These results are heavily dependent on the quality of the HEXTE background subtraction at the highest energies, but provide good support for our supposition that a non-thermal Comptonisation component starts to become important in the intermediate state and affects the inferred location of the cutoff for the thermal electron distribution.

The relative strength of the harmonic to the QPO is dependent on the optical depth, with lower optical depths producing stronger harmonics (see \cite{Axelsson2013a} and references within). Such an effect can be seen in the reproduced QPO waveforms, with two maxima seen per cycle (see top panel of Fig.~\ref{fig:waveform}). As the source softens, and the flow becomes optically thick, the harmonic becomes weaker, leading to a waveform with a single maximum. The maximum of this waveform occurs when the disk at the precession phase that is closest to face-on for the observer (see bottom panel of Fig.~\ref{fig:waveform}).

\subsubsection{Is this a reasonable value for $\tau$?}

The presence of the hard X-ray lags \citep{Priedhorsky1979} can indicate a temperature gradient in the corona in the context of the propagating fluctuations model \citep{Lyubarskii1997, Kotov2001} In this case, the viscous propagation of the material as it moves inwards through regions of increasing temperature, leads to the variability at lower energies preceding those at higher energies. 

Spectral fitting to obtain optical depths can often yield poor fits for medium signal-to-noise ratio with high $\tau$, especially for a multi- temperature black body. Here, we make an estimate of the optical depth of the corona for a simple model. Consider a torus with a major radius $R$ and minor radius $R/2$ (i.e. an aspect ratio of 2). The density of the torus can be approximated to be 

\begin{equation}
    \rho = \frac{\dot{M}t_{visc}}{\pi^2 R^3}
\end{equation}

where $\dot{M}$ is the mass accretion rate and the viscous timescale $t_{visc} \sim R^2/\alpha c_s H $. $H$ is the scale height with $H/R \sim$ 1 for a torus, and $\alpha$ is the viscosity parameter \citep{Shakura1969}. The sound speed $c_s$ is given by $\sqrt{kT/m_p}$ where $m_p$ is the proton mass. Integrating the density along the radius yields the surface density 

\begin{equation}
    \Sigma = \int_{R_{in}}^{R_{out}} \! \frac{\dot{M}R}{\pi^2 \alpha c_s R^3} \, \mathrm{d}R \sim \frac{\dot{M}}{\pi^2 \alpha c_s} [\frac{1}{R_{in}}-\frac{1}{R_{out}}]
\end{equation}

The optical depth of this torus is then given by 

\begin{equation}
    \tau = \sigma N =\sigma \Sigma / m_p
\end{equation}

The high energy cut off during the intermediate states has been measured to be kT $\sim$ 130 keV \citep{Motta2009}. Using a value of $\alpha$ = 0.1 , $R = 10^7$m and a value of ${\dot{M}}$ between $10^{18}$ and $10^{19}$ g/s gives a value of $\tau$ between 1.5 and 15. This agrees with the estimated $\tau$ required to explain the scenario described in Section~\ref{sec:physical}, where at low $\dot{M}$ values $\tau$ < 3.2 and at $\tau$ increases at higher accretion rates causing the smearing out of the high frequency variations.

\section{Conclusions}
\label{sec:conclusion}

We have shown that in GX 339-4, in both type B and C QPOs the fundamental frequency is coupled to higher harmonics. Additionally, type C QPOs show coupling between QPO frequency and the broadband noise, while type B QPOs do not.  We have also found that the bicoherence gradually changes over the duration of the X-ray outburst from a `web' pattern to a `hypotenuse' pattern.

We also reconstruct the QPO waveforms for 6 observations from the values of the phase difference between harmonics. This was obtained using the value of the biphase, which contains information about the underlying lightcurve. We find that the phase difference evolves, leading to a change in the QPO waveform.

Finally, we present the scenario of a moderate increase in the optical depth of the hard X-ray emitting corona to explain the gradual change in the bicoherence pattern that is seen as the source moves from a hard intermediate to a soft intermediate state.

\section*{Acknowledgements}
The authors would like to thank Abbie Stevens, Adam Ingram, and Mariano Mendez for interesting discussions. We also thank the anonymous referee for comments that improved the clarity of the paper.



\bibliographystyle{mnras}
\bibliography{Bicoherence} 




\bsp	
\label{lastpage}
\end{document}